\documentclass[aps,prb,superscriptaddress,nobibnotes,notitlepage,twocolumn,preprintnumbers,10pt]{revtex4-1}
\usepackage[utf8x]{inputenc}
\usepackage{graphicx}
\usepackage{fancyvrb}
\usepackage{amsmath}
\usepackage{mathrsfs}
\usepackage{amsfonts}
\usepackage{ulem}
\usepackage[usenames,dvipsnames]{color}
\usepackage{url}
\usepackage[colorlinks=true,linkcolor=blue]{hyperref}
\usepackage{textcomp}
\usepackage{subfig}
\usepackage{float}
\usepackage{miller}

\graphicspath{{./Fig/}}

\begin{document}
\title{Thermally driven domain wall motion in Fe on W\hkl(110)}
\author{Jonathan Chico}
\author{Corina Etz}
\affiliation{Uppsala University, Department of Physics and Astronomy, Materials Theory Division \\ Box 516 SE-751 20 Uppsala Sweden.} 
\author{Lars Bergqvist}
\affiliation{KTH Royal Institute of Technology, Dept. of Materials and Nano Physics \\ Electrum 229, SE-164 40 Kista, Sweden}
\author{Olle Eriksson}
\affiliation{Uppsala University, Department of Physics and Astronomy, Materials Theory Division \\ Box 516 SE-751 20 Uppsala Sweden.}
\author{Jonas Fransson}
\affiliation{Uppsala University, Department of Physics and Astronomy, Materials Theory Division \\ Box 516 SE-751 20 Uppsala Sweden.}
\author{Anna Delin}
\affiliation{Uppsala University, Department of Physics and Astronomy, Materials Theory Division \\ Box 516 SE-751 20 Uppsala Sweden.}
\affiliation{KTH Royal Institute of Technology, Dept. of Materials and Nano Physics \\ Electrum 229, SE-164 40 Kista, Sweden}

\author{Anders Bergman}
\affiliation{Uppsala University, Department of Physics and Astronomy, Materials Theory Division \\ Box 516 SE-751 20 Uppsala Sweden.} 

\begin{abstract}
It has recently been shown that domain walls in ferromagnets can be moved in the presence of thermal gradients. In this work we study the motion of narrow domain walls in low-dimensional systems when subjected to thermal gradients. The system chosen is a monolayer of Fe on W\hkl(110) which is known to exhibit a large anisotropy while having a soft exchange, resulting in a very narrow domain wall. The study is performed by means of atomistic spin dynamics simulations coupled to first-principles calculations. By subjecting the systems to a thermal gradient we observe a temperature dependent movement of the domain wall as well as changes of the spatial magnetization profile of the system. The thermal gradient always makes the domain wall move towards the hotter region of the sample with a velocity proportional to the gradient.
The material specific study is complemented by model simulations to discern the interplay between the thermal gradient, magnetic anisotropy and the exchange interaction, and shows that the larger DW velocities are found for materials with low magnetic anisotropy. The relatively slow DW motion of the Fe/W\hkl(110) system is hence primarily caused by its large magnetic anisotropy.
\end{abstract}

\date{\today}
\maketitle

\section{Introduction}
The recently discovered spin Seebeck effect (SSE) \cite{seebeck_nat} shows that a magnetic material subjected to a thermal gradient will develop a\textquotedblleft spin voltage\textquotedblright, which causes an accumulation of spins of one character at one edge of the sample and an accumulation of spins of the opposite character on the other. Although originally observed in a conductor the effect has also been observed in magnetic semi-conductors \cite{seebeck_semi} and insulators \cite{seebeck_in}. This discovery has opened new research paths due to it's possible uses in spintronic devices. The mechanism of the SSE is not completely understood due to the wide variety of materials which have presented the effect, a magnonic contribution has been proposed \cite{nowak}, although a more complete description can be obtained by considering a phonon driven contribution \cite{phonon}. 

Domain walls (DW) have been thoroughly studied in the last decades due to their importance for practical applications, above all as data storage devices, therefore their controlled movement under magnetic fields and spin polarized electronic currents (from voltage or temperature differences between the edges of the samples) are the subject of intense research~ \cite{bauer2010,Sampaio2012,Xia2010,Gerrit2010,Stiles2008,Li2004, Niebala2007}, purely magnonic DW movement has also been studied by using selective spin wave excitations \cite{Wang2012,Yan2011}. By introducing thermal gradients, recent studies \cite{nowak,klaui1,klaui2} have shown the possibility to move a domain wall through thermally excited magnons both experimentally and theoretically. 

Hitherto, most studies on the manipulation of domain walls have been performed on bulk materials and in an regime where domain walls are typically quite wide. In this work we focus on thermally induced domain wall motion in the limit of a single monolayer. In addition we focus on a well characterized system, a monolayer of Fe on W\hkl(110), which is known to have a large magneto crystalline anisotropy and a softened magnetic exchange coupling which result in a system with very narrow domain walls. On this system we study numerically the dynamics of domain walls in the presence of thermal gradients. Since the domain walls for the system are very narrow, a method with high spatial resolution is needed and thus we here perform atomistic spin dynamics simulations\cite{antropov,skubic} coupled to ab initio calculations. Since the current system is metallic, a real temperature gradient would infer an electronic transport due to the Seebeck effect which in turn would affect the magnetization dynamics of the system 
through spin transfer torque effects. These effects are however not considered in this study as we are interested in the pure magnonic effects.

The rest of this paper is organized as follows. First we introduce the theoretical framework for the simulations in Sec.~\ref{sec:method}. Then the material specific study of the domain wall motion in Fe/W\hkl(110) is reported in Sec.~\ref{sec:few}. In order to obtain a more complete understanding of the phenomenon of thermally induced domain wall motion, specifically the role of the the strength of the magnetic exchange and anisotropy energies, we perform a number of model simulations where the ratio between exchange and anisotropy is varied for a 2D monolayer. These model simulations are presented in Sec.~\ref{sec:model}. We end the paper by analyzing the results and discussing future developments in the field.

\section{Method}
\label{sec:method}
In order to investigate the dynamics of the DW introduced in the system, we perform atomistic spin dynamics\cite{antropov} simulations implemented in the UppASD package \cite{skubic,hellsvik}. Each atomic moment is considered to be a three dimensional (3D) vector with an equation of motion given by the expression:
\begin{equation}
 \frac{\partial \mathbf{m_i}}{\partial t}=-\gamma \mathbf{m_i} \times \mathbf{B}^{eff}_i-\gamma \frac{\alpha}{m}\left[\mathbf{m_i} \times \left[ \mathbf{m_i} \times \mathbf{B}^{eff}_i\right]\right]
 \label{eq:LL}
\end{equation}
where the effective field is given by: 
\begin{equation*}
 \mathbf{B}^{eff}_i=\underbrace{\mathbf{B}_i}_\text{Hamiltonian}+ \underbrace{\mathbf{b}_i(T)}_\text{Stochastic}
\end{equation*}
the $\mathbf{B}_i$ field takes into account the interactions in the system and is defined as:
\begin{equation*}
 \mathbf{B}_i=-\frac{\partial \mathcal{H}}{\partial \mathbf{m}_i}
\end{equation*}
while the $\mathbf{b}_i(T)$ stochastic field is included in order to account for temperature effects by using Langevin dynamics~\cite{skubic}.

\begin{figure}
 \centering
 \includegraphics[width=0.5\textwidth]{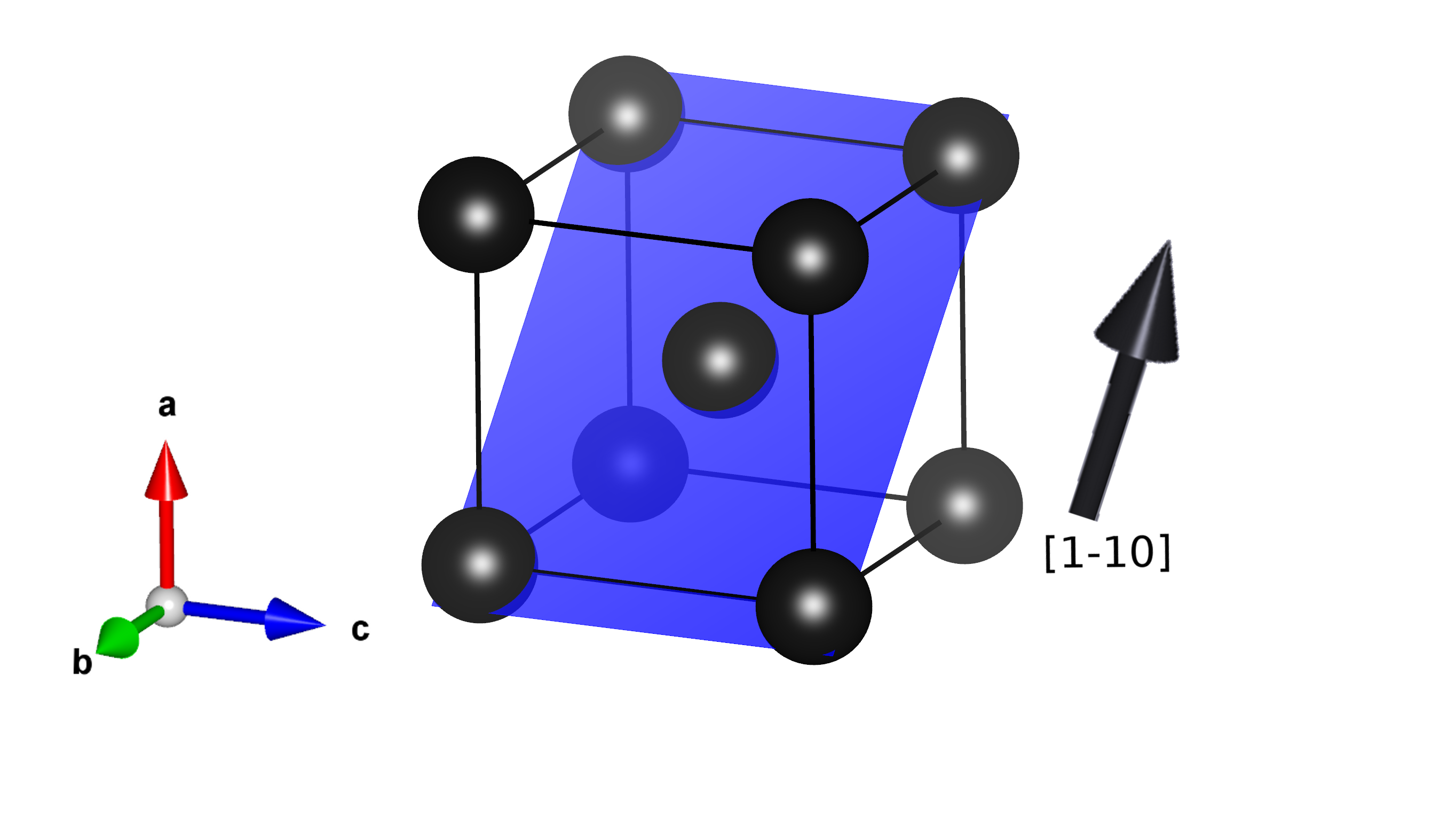}
  \caption{Unit cell structure and easy axis of Fe on W\hkl(110)~\cite{VESTA}, with a lattice parameter of 3.16 \AA. The direction \hkl[1 -1 0] denotes the magnetization easy axis.}
  \label{fig:Figure1}
\end{figure}

The Hamiltonian used in this approach is an extended Heisenberg Hamiltonian, which for the present study only includes two terms, accounting for the exchange interactions and (uniaxial) anisotropy: 
\begin{equation}
 \mathcal{H}=\underbrace{-\frac{1}{2}\sum_{i\neq j} J_{ij} \mathbf{m}_i \cdot \mathbf{m}_j}_\text{exchange}+\underbrace{K\sum_{i}\left( \mathbf{m}_i\cdot \mathbf{e}_k\right)^2}_\text{uniaxial anisotropy}
 \label{eq:hamiltonian}
\end{equation}
In Fe on W\hkl(110) the easy magnetization axis points along the x-axis of the simulation cell which corresponds to the \hkl[1 -1 0] crystallographic axis (Fig.~\ref{fig:Figure1}). The exchange interactions $J_{ij}\textrm{'s}$ and magnetic moments used were obtained from ab initio calculations done in a previous study~\cite{anders}. The $J_{ij}\textrm{'s}$ were considered up to the 40th neighbor shell in order to properly account for long-range effects. To each Fe atom corresponds a calculated moment of $m=2.13 \mu_B$. The value of the anisotropy constant is $K=0.338\textrm{ mRyd}$ and the magnetization easy axis direction is the \hkl[1-10] as reported in an experimental study~\cite{Wiesendanger2001}. This system presents a very large anisotropy while having a soft exchange \cite{anders} which is expected to lead to a very narrow domain wall. Since the size of the systems considered in this study is very limited, dipolar interactions are not important for neither the magnetic ground state nor the dynamical 
properties and therefore we have not included dipolar interactions in our simulations. The domain walls present in this study are thus metastable and determined only by the competition between Heisenberg exchange interactions and the magneto crystalline anisotropy. 
\\In this study we have only considered scalar Heisenberg exchange. It is known that in Fe on W\hkl(110) systems also chiral Dzyaloshinskii-Moriya interactions (DMI) are present\cite{Zakeri2010}. The presence of DMI could introduce a chirality to the domain wall which might effect the dynamics of the DW. While these effects are out of the scope for this work, they could be very interesting do address in a related future study.
\\ 
In these atomistic spin dynamics simulations the damping parameter $\alpha$ governs the rate of the dynamics. While this parameter can be measured and even estimated by ab initio calculation we are not aware of a reported value of the damping for this system and for this reason we set the value to $0.1$ which is of similar scale to what can be expected for Fe related systems in low dimensions\cite{Steiauf2005}.

\section{F\lowercase{e} on W \hkl(110) Simulations}
\label{sec:few}
\begin{figure}
 \centering 
 \subfloat[Domain wall before equilibration]{\label{fig:Figure2a}\includegraphics[width=0.35\textwidth,height=2cm]{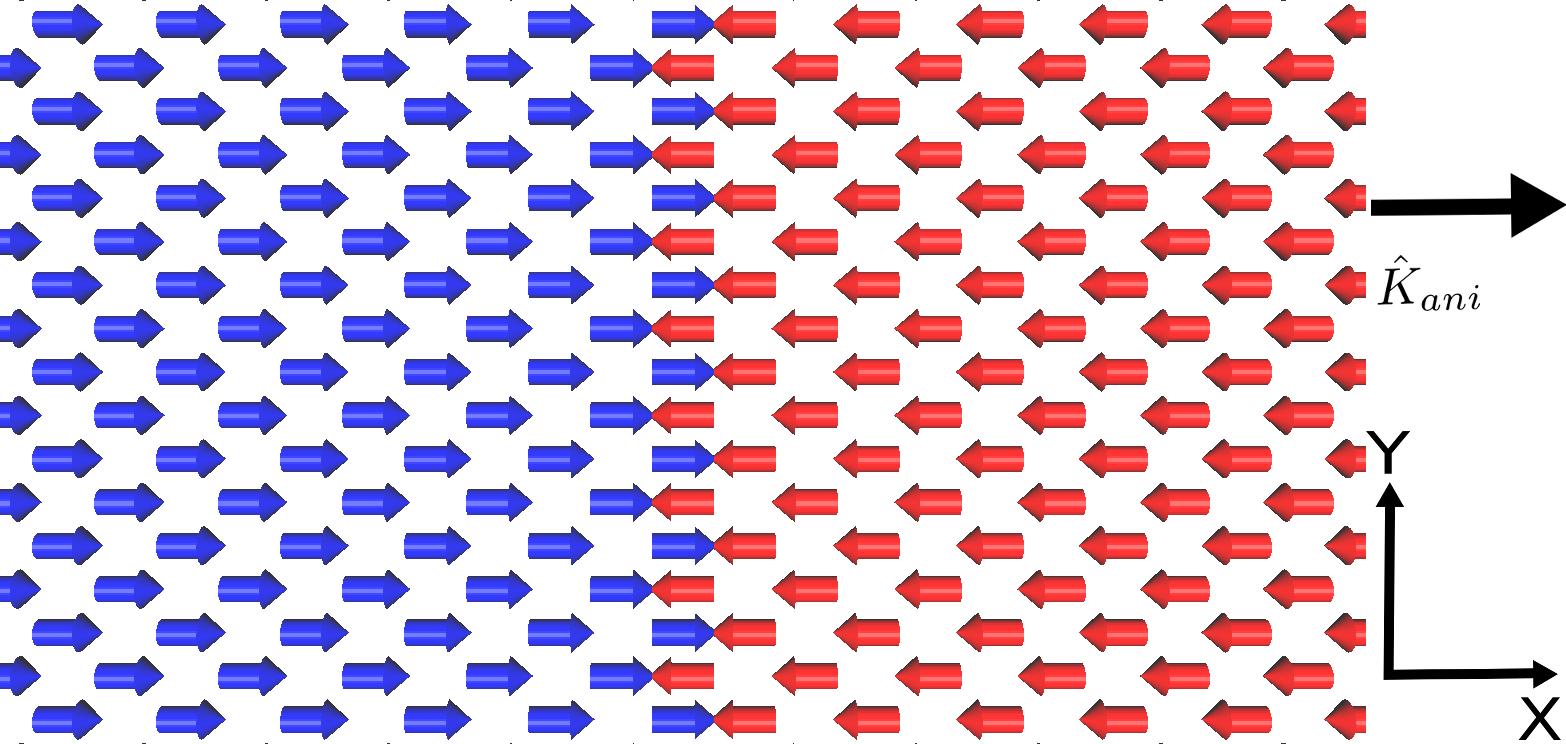}}\\
  \vspace{-3mm}
 \subfloat[Domain wall equilibrated at T=20K]{\label{fig:Figure2b}\includegraphics[width=0.35\textwidth,height=2cm]{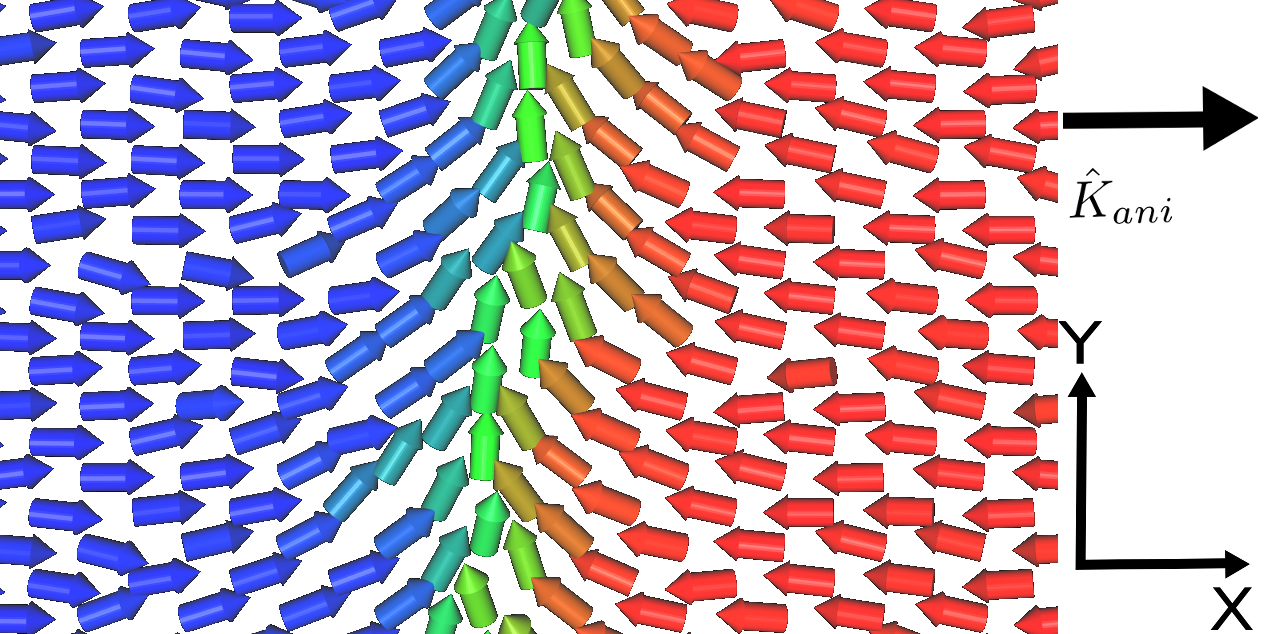}}
 \caption{Visualization of the considered domain wall, before and after thermal equilibration. The topmost figure shows how the domain wall is introduced in the system by flipping half of the spins in the system. The lower figure shows a snapshot of the same domain wall after being equilibrated at a temperature of 20 K.
 }
 \label{fig:Figure2}
\end{figure}

In this study we consider a geometry constructed by 100 repetitions along the \hkl[1 -1 0] direction (this axis will be referred to as the x axis from now on) and 40 repetitions in the \hkl[1 0 0] (y axis). In this configuration the easy magnetization axis is parallel to the long axis of the simulation box (Fig.~\ref{fig:Figure2a}). The system is described with free boundary conditions. It is worthwhile to emphasize that the inter-atomic distance along the x-axis is $\sqrt{2}a$ with $a=3.16$\AA, while along the y-axis the distance is just $a$. 

The ground state of a monolayer of Fe/W\hkl(110) is ferromagnetic, with all the atomic magnetic moments aligned along the $x$ direction. Therefore to be able to introduce a domain wall in this system half of the spins in each sample are flipped (as show in figure ~\ref{fig:Figure2}) in the simulations. This means that for the initial state the magnetization of the samples is zero. 

It is known that the DW width in a material at $T=0K$ is given by the ratio between the exchange stiffness $A$ and the uniaxial anisotropy constant $K$, so that $\delta=\pi\sqrt{\frac{A}{K}}$~\cite{nowak,oguchi, Chantrell2008}. In the initial state the DW is infinitely sharp. To obtain a DW with a width which takes into account the system parameters and the temperature effects, the system is equilibrated at a constant temperature, $T_{ave}$, which gives a finite DW width Fig.(~\ref{fig:Figure2b}). This width depends on the equilibration temperature, and as this temperature approaches zero the DW width is $\sim 3 \textrm{ nm}$. Due to the narrow width of the DW the maximum temperature in which it is still stable, $T_h$, can differ greatly from the $T_C$ of the material \cite{Kazantseva}. Note also that the domain wall in Fig.(\ref{fig:Figure2b}) is slightly skewed. Averages over several simulations removes this skewness.  

Diverse methods have been implemented to try to move the domain wall from its equilibrium position in a controlled way. One of the most used techniques in controlled DW movement, is the phenomena that is the result of  the application of a spin polarized current over a region of non homogeneous magnetization. The polarized current exerts a torque over the magnetic moments which is known as the Spin Transfer Torque (STT), this angular momentum transfer can be used to move domain walls \cite{Stiles2008,Li2004, Niebala2007}. The strength of the torque acting over the magnetic texture depends on the magnitude of the current density applied to the sample. A thermal gradient can also induce spin polarized thermoelectric currents in ferromagnetic metals. The Seebeck effect describes the creation of an electric current when a metal is subjected to a thermal gradient, this has led to the investigation of thermally induced STT's. This effect has been studied recently in great detail \cite{bauer2010,Sampaio2012,Xia2010,
Gerrit2010}, using linear response formalism. The parameter that controls the strength of the torque is calculated to be given by $\displaystyle \tau \propto \frac{\Delta T}{T_{ave}}$ \cite{bauer2010,Wees}. These effects are not considered in this work, as no torque induced by thermoelectric effects are considered in our simulations, only the magnonic effects resulting from the application of a thermal gradient over the magnetic sample are taken into account.

\begin{figure}
 \centering 
 \includegraphics[width=0.4\textwidth]{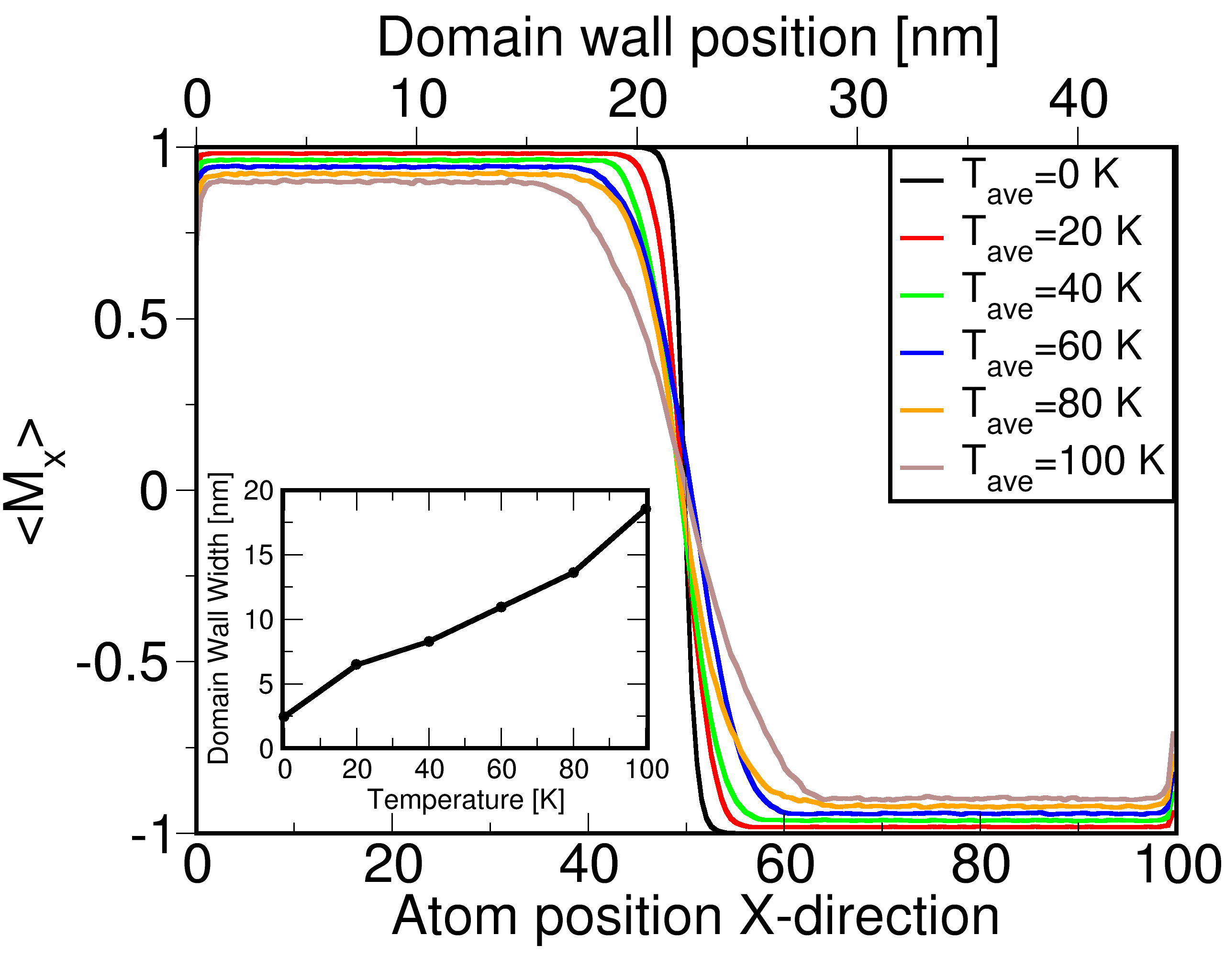}
 \caption{Magnetization profile along the long axis of the sample, the lower axis allows us to see the extension of the DW in cell repetition units while the upper axis shows it in nanometers. The inset show a linear dependence of DW width with temperature.}
 \label{fig:Figure3}
\end{figure}

The DW movement can be tracked by analyzing the magnetization component projected on the x-axis which in both samples allows us to evaluate the relative sizes of the domains. We find that the average magnetization along the x-axis during the equilibration exhibits small fluctuations around zero due to thermal effects. In general, the DW does however remain at its initial position, in the middle of the sample. Figure~\ref{fig:Figure3} shows the profile of the domain wall for equilibration temperatures $T_{ave}$ ranging from $0\textrm{K}$ to $100\textrm{K}$. On the edges of the simulation box the magnetization is saturated, while in the middle there is a smooth change in the moments direction which shows the profile of the domain wall.

It is possible to see that as the temperature increases the domain wall width increases. This is due to the thermal fluctuations, which adds energy to the system and thereby allows a broadening of the inhomogeneous magnetization region that determines the domain wall. We find that the DW width increases linearly with an increase in temperature (see insets in Fig.~\ref{fig:Figure3}).

After equilibration, the system is subjected to a uniform thermal gradient chosen such that the local temperature of the DW (and the average temperature of the system) does not change and the thermal energy of the sample remains constant. The gradient is given by:
\begin{equation}
 T({\it l})=T_{min}+\frac{\Delta T}{\chi_{max}}\cdot {\it l}
 \label{eq:gradient}
\end{equation}
with $T_{min}=T_{ave}-\frac{\Delta T}{2}$, and $T_{max}=T_{ave}+\frac{\Delta T}{2}$ and ${\it l}$ is the position in the sample. Simulations were performed for $T_{ave}=20\textrm{K}$ to $T_{ave}=100\textrm{K}$ with gradients going from $\Delta T=1\textrm{K}$ to $\Delta T=50\textrm{K}$. 

\begin{figure}
 \centering
  \includegraphics[width=0.4\textwidth]{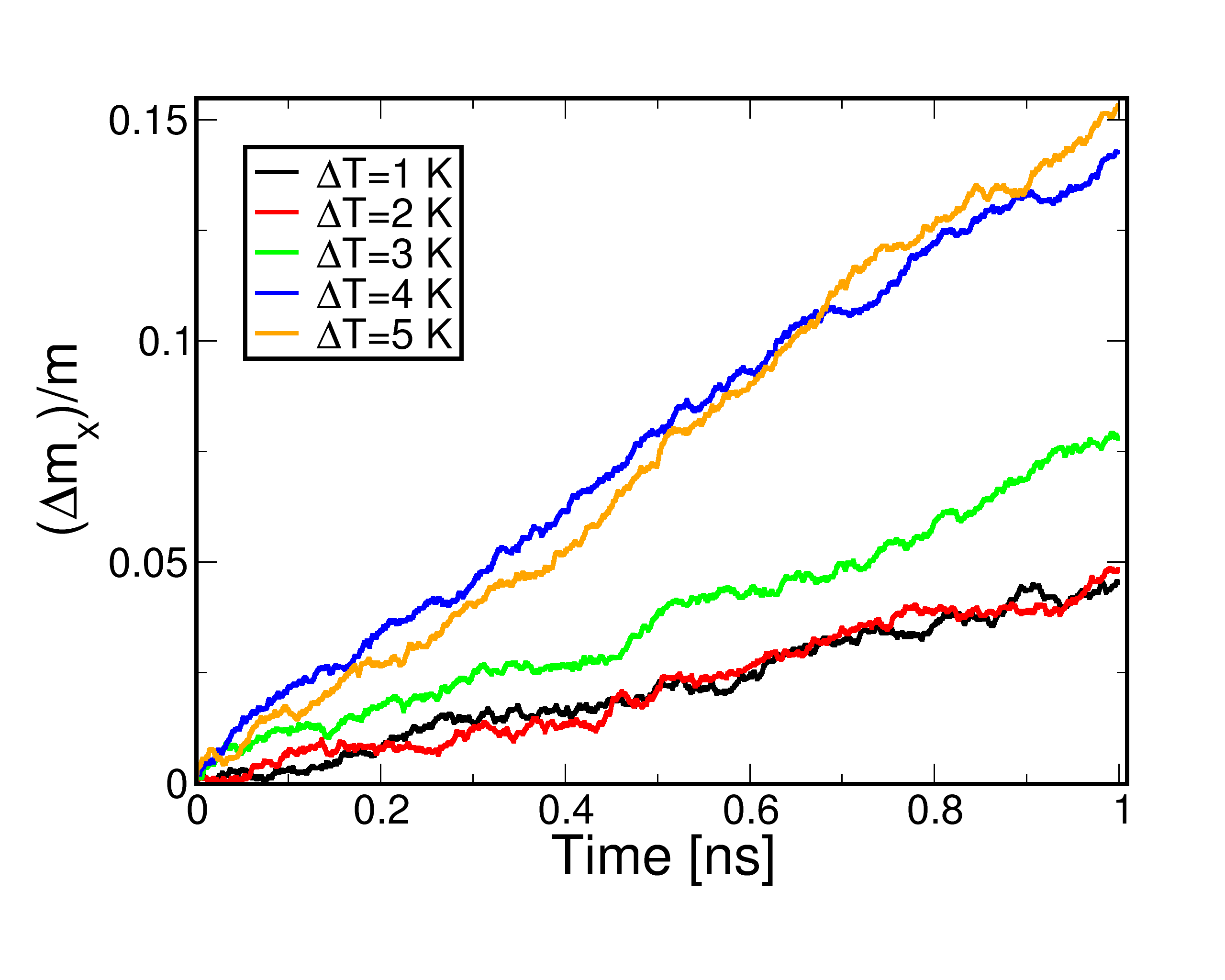}  
  \caption{$\frac{\Delta\langle m_x\rangle}{m}$ as a function of time for $T_{ave}=20\textrm{K}$, where $\Delta \langle m_x\rangle=m_x(0)-m_x(t)$, where $m_x(0)$ is the magnetization of the system just before the thermal gradient is introduced. As the strength of the gradient increases it is possible to see that the magnetization increases more rapidly, which indicates an increase in the DW speed.}
  \label{fig:Figure4}
\end{figure}

Our main finding is that the thermal gradient makes the DW move towards the hotter region of the sample, which is consistent with what has been previously observed experimentally\cite{Torrejon2012} and in other numerical simulations\cite{nowak}. Figure~\ref{fig:Figure4} shows how the normalized average magnetization in the x-direction increases as a function of time when the DW is subjected to a thermal gradient. This carries information on the DW displacement in time, since when $\langle m_x \rangle=0$ the sample has two equally large domains with up and down spins, whereas a finite magnetization implies that one domain has grown on behalf of the other, i.e. the DW has moved. The slope of the curve in Fig.~\ref{fig:Figure4} depends on the strength of the gradient, becoming steeper with an increased temperature gradient, which means that the DW moves faster towards the hotter end of the sample. 

The movement of the DW towards the hotter region due to the thermal gradient is a finding that can be interpreted in several ways. It has been previously argued that the domain wall moves as it is dragged by thermally excited magnons\cite{nowak}. A magnonic spin current has been demonstrated to be able to move DWs \cite{Yan2011,Wang2012}, due to the magnonic Spin Transfer Torque. It is also important to notice that due to the size of our sample and for the temperatures taken into consideration, the ground state of the system is ferromagnetic. In larger systems, where the dipolar interaction becomes dominant the formation of magnetic domains minimizes the energy, making the ground state a domain structure. However the interactions in our system do not favor the existence of a DW. This means that under this circumstances the DW is a metastable state and therefore this allows us to think of the DW under the thermal gradient as an ordering process. 

Temperature enters in Eqn.~\ref{eq:LL} as stochastic field $\textbf{b}_i(T)$, in the equilibration step the temperature is constant, therefore $\textbf{b}_i(T)$ is not biased, the strength of the fluctuations is the same over the whole sample, and therefore can't destabilize the DW and cause its motion. It is important to note that for sufficiently high temperatures the domain wall can indeed move even in a constant-temperature setting but then the movement would be a completely unbiased Brownian motion resulting in a random walk with no net displacement on average. 
However when a temperature gradient is added it creates a bias in the stochastic field. The bias facilitates domain wall movement towards the hotter side of the sample while it makes it less likely for the domain wall to move in the direction of the cold end of the sample. 

\begin{figure}
 \centering
 \subfloat[DW speeds for $\frac{T_{max}}{T_{ave}}\sim1$]{\includegraphics[width=0.35\textwidth]{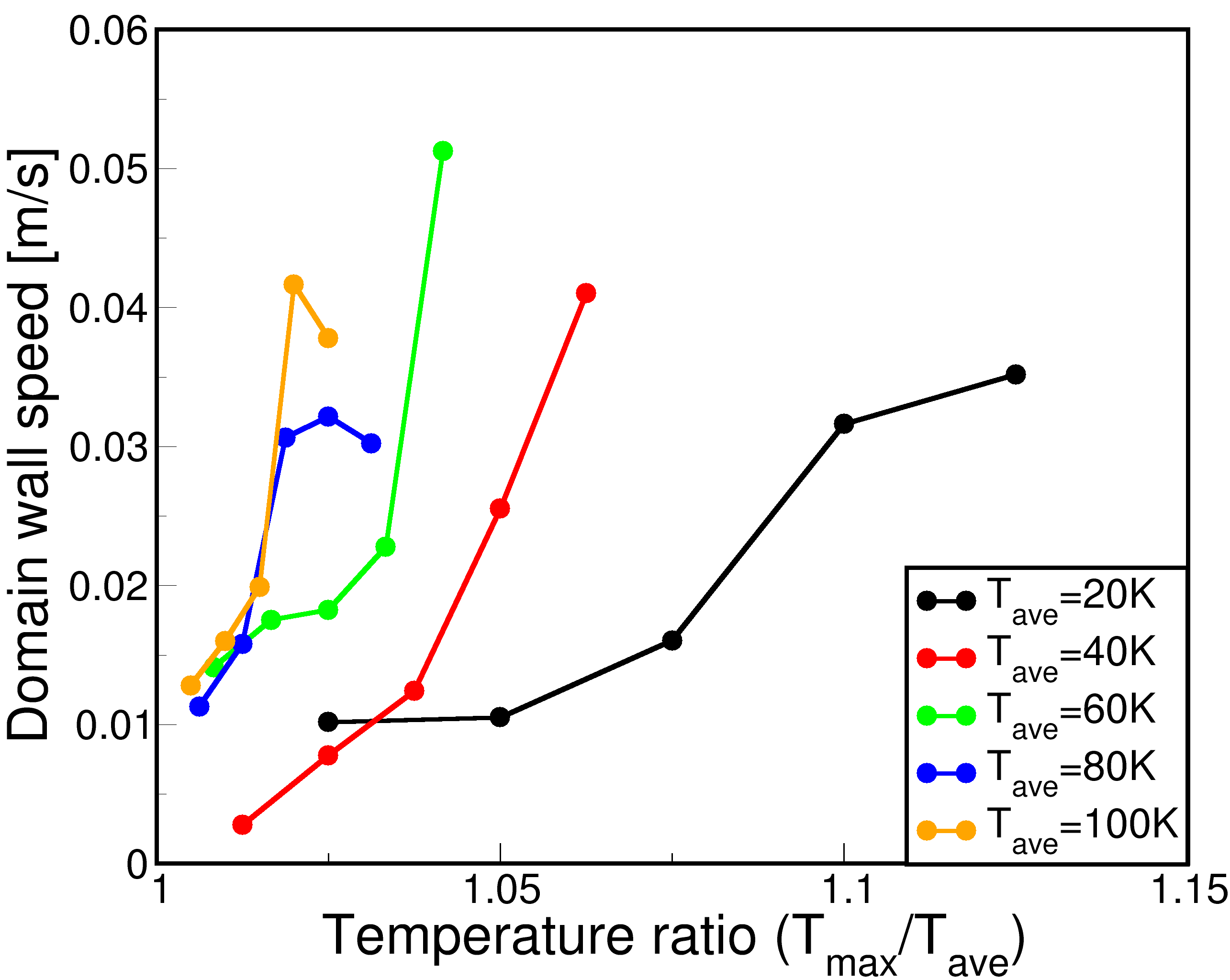}}\\
  \subfloat[DW speeds for $\frac{T_{max}}{T_{ave}}\gg1$]{\includegraphics[width=0.35\textwidth]{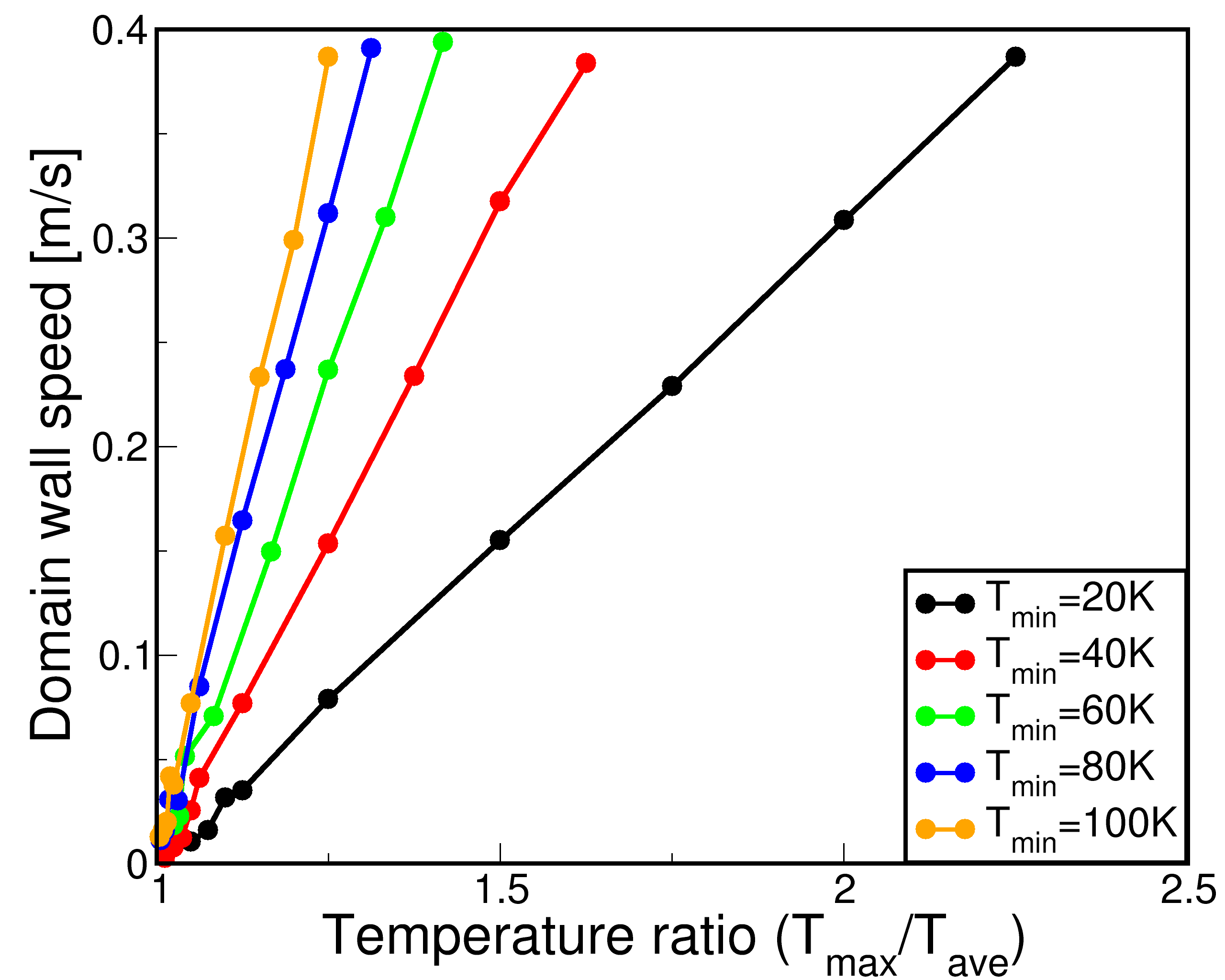}}
 \caption{DW speed as a function of $\frac{T_{max}}{T_{ave}}$ for fixed $T_{ave}$. As can be seen, the $T_{ave}$ influences the attainable speed for a given gradient. When $T_{max}$ is much larger than $T_{ave}$ the DW speed depends linearly on the temperature ratio, if $T_{max}\sim T_{ave}$ the bias in the thermal fluctuations is not strong enough which causes the dependence to be non-linear.}
 \label{fig:Figure5}
\end{figure}

We have studied the DW speed in the system for various values of $T_{ave}$, while varying the gradient for each choice of $T_{ave}$. By changing the temperature gradient, and implicitly the $\frac{T_{max}}{T_{ave}}$ ratio, we can influence the domain wall speed, as can be seen in Figure~\ref{fig:Figure5}. Moreover, if $T_{max}$ is comparable with $T_{ave}$, the relationship between the DW speed and the temperature ratio is observed to be non-linear. If $\frac{T_{max}}{T_{ave}}$ is just above one, the thermal fluctuations of atomic spin on either side of the DW have similar strength, even if the atomic moments closer to the hotter edge will have stronger fluctuations this is not strong enough to completely dominate the time evolution of the magnetization. The competition between the \textquotedblleft noise\textquotedblright or unbiased part of the thermally induced fluctuations, and the \textquotedblleft signal\textquotedblright or the biased part of these fluctuations can lead to a lower speed of the DW, 
if the bias is not strong enough to overcome the random fluctuations which tend to push the DW in random directions. This is observed in situations where $T_{ave}$ is high, as strong thermal fluctuations make the DW fluctuate around its equilibrium position. In these situations a small increase in the thermal gradient strength might lead to a decrease in the DW speed. It also can be noticed that after a certain threshold ratio when $T_{max}$ is much larger than $T_{min}$ the DW speed increases linearly with an increasing temperature gradient.

The domain wall speeds obtained for this system are much lower than the ones reported for DW motion driven by a STT\cite{Li2004}. There could be several collaborating reasons for this behavior. The main reason for this is the very narrow DW width itself that our sample exhibits. A broader domain wall will obviously have a larger temperature difference between the end points than would a narrow domain wall at the same temperature gradient. A narrow domain wall also infers that the magnetic anisotropy of the system is strong, as in the present case, which makes the DW harder to move thus requiring a larger gradient. 

\section{\label{sec:model}Model system Simulations}

As the DW speed obtained in this study is much smaller than what has been previously reported\cite{nowak,Li2004}, we studied a model system to try to determine which is the source of this behavior, and also to get a more complete picture of the physics of this effect. One of the most important aspects of the Fe on W\hkl(110) system is that the domain walls that can be introduced are very narrow. To test if this aspect is determinant to the DW speed, model calculations in which the value of the anisotropy constant is varied were performed. A change in the anisotropy constant $K$, as mentioned before, should result in a change of the DW width. This variation of the DW width allows us to achieve a qualitative behavior of the DW when it is subjected to a thermal gradient and an understanding of which configurations could be more convenient for technological applications.

For the model calculations several simplifications were done, the unit cell considered is a square lattice with a lattice parameter of $a=4.5$\AA \hspace{1pt} with $100\times40$ repetitions; this was chosen such that the length of the model simulations were comparable to the ones of the material specific calculations. The Hamiltonian used is the one described in eq.~\ref{eq:hamiltonian}. In this case the exchange interactions are considered only between nearest neighbors and their strength is considered to be $J_{ij}=1\textrm{ mRyd}$. Dipolar interactions were neglected here as in the material specific case. The magnetic moments were set to $m_i=1 \mu_B$.  
The easy magnetization axis is along the x-direction and the anisotropy constant $K$ is varied between $K=0.1\textrm{ mRyd}$ and $K=1.0\textrm{ mRyd}$. In this way it was possible to observe the effect that the different $\frac{J}{K}$ ratios would have over the thermally assisted DW motion.

\begin{figure}
 \centering
 \includegraphics[width=\columnwidth]{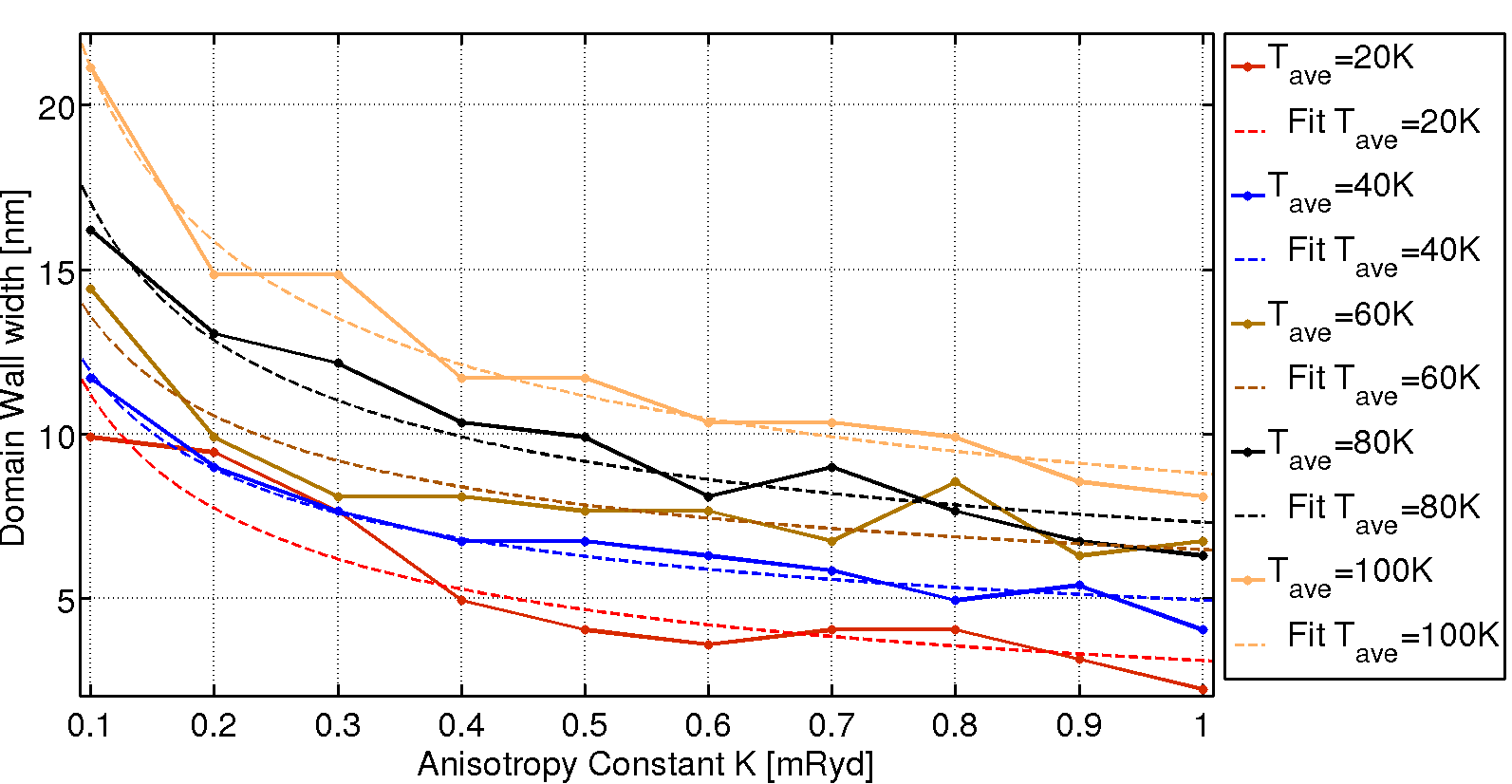}
 \caption{Domain wall width variations induced by the changes in anisotropy for different equilibration temperatures $T_{ave}$.}
 \label{fig:Figure6}
\end{figure}

Samples with the same exchange but different anisotropy strength were studied. As in the case of Fe on W\hkl(110) the systems were equilibrated at different temperatures, $T_{ave}$. As can be seen in  Figure~\ref{fig:Figure6}, as the anisotropy constant decreases the DW width increases. It is also worth noticing that if one considers a fixed anisotropy and increase the equilibration temperature the DW thickness also increases, this is consistent with what was observed in the insets in Figure~\ref{fig:Figure3}, where the DW width increases linearly with temperature.

As in the material specific case, discussed in the previous section, the magnetization profile allows us to see that the DW is located at the middle of the sample. This means that the random fluctuations generated by the stochastic field in the equilibration process are not strong enough to destabilize the DW even when the value of the anisotropy is lowered. In Figure~\ref{fig:Figure6} it is possible to observe that for a fixed $T_{ave}$ as the anisotropy increases the DW width decreases in a non linear fashion. For $T_{ave}=0K$ the width is known to follow $\propto \frac{1}{\sqrt{K}}$, hence the width obtained from all the numerical simulations was fitted to the expression $\frac{a_1}{\sqrt{K}}+a_0$, where $a_1$ and $a_0$ are fitting parameters. The agreement within the fitting and the data is quite good (see Fig.~\ref{fig:Figure6}), showing some deviation which results from the fluctuations caused by finite temperature effects.

After the equilibration, in order to make the DW move towards the edge of the sample, we subject each system to a linear thermal gradient described in Eq.~\ref{eq:gradient}. 
\begin{figure}
 \centering
 \subfloat[$T_{ave}=20K$ and $K=0.1 \textrm{ mRyd}$]{\label{fig:Figure7a}\includegraphics[width=0.35\textwidth]{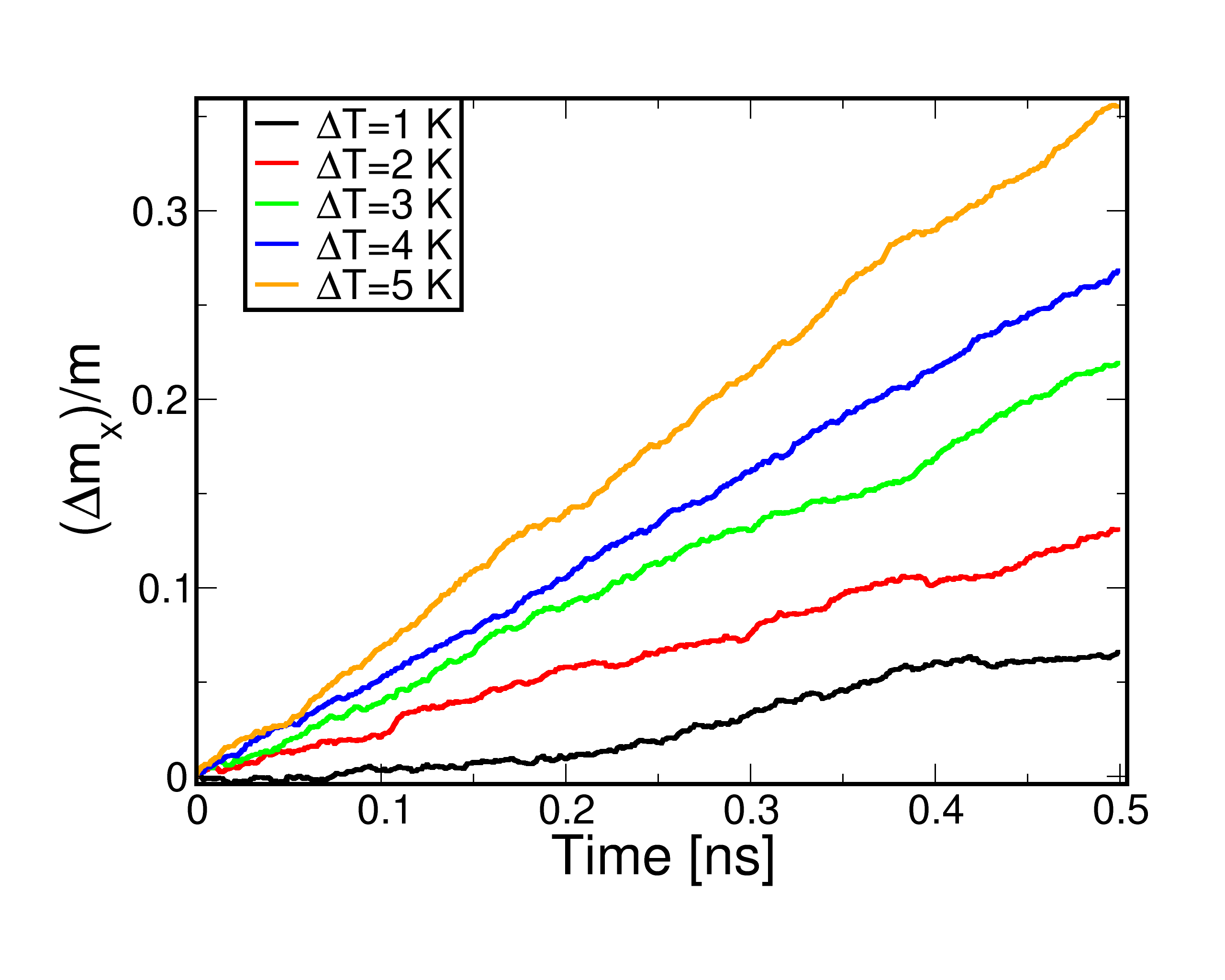}}\\ 
 \subfloat[$T_{ave}=20K$ and $K=1 \textrm{ mRyd}$]{\label{fig:Figure7b}\includegraphics[width=0.35\textwidth]{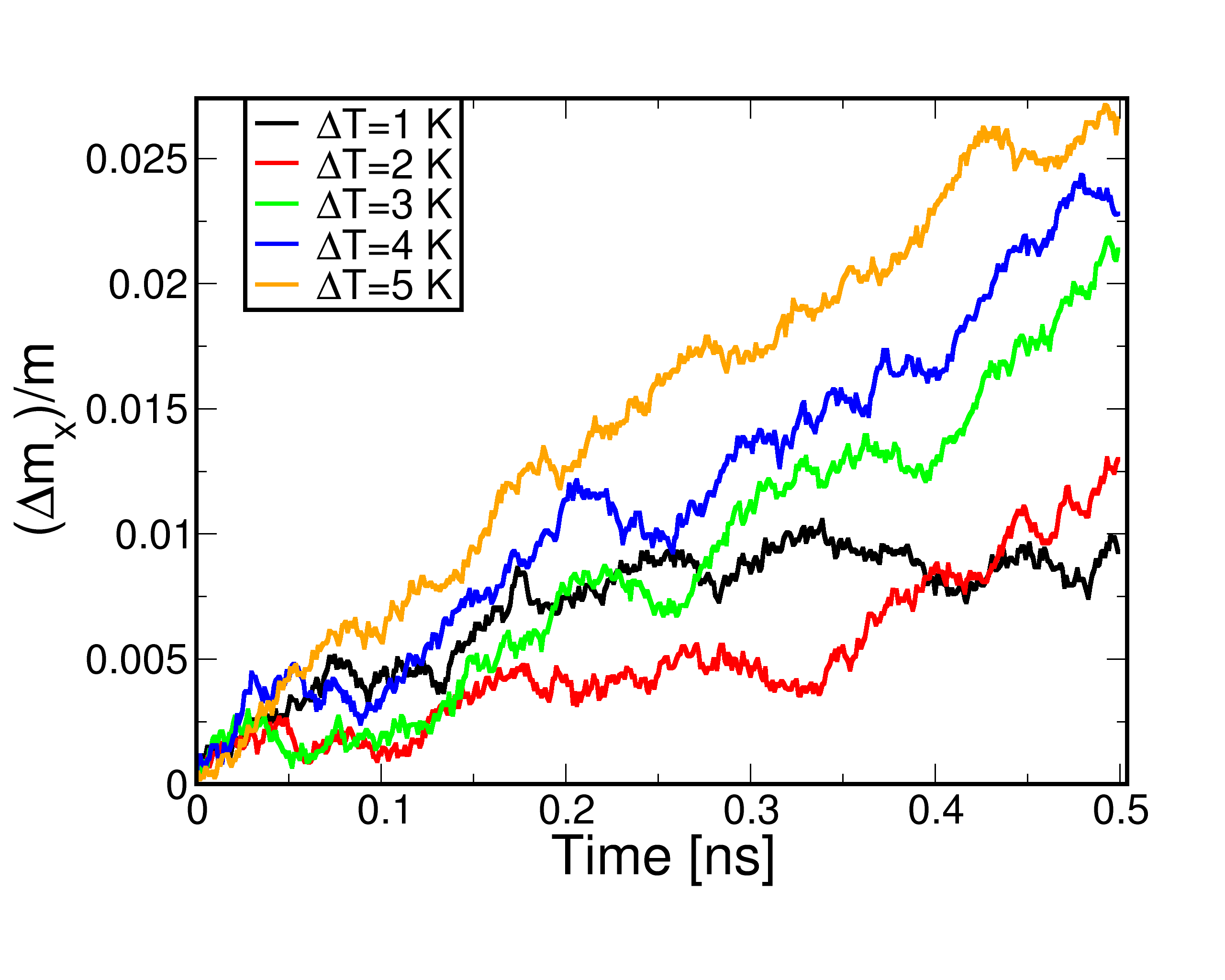}}
 \caption{\small{Difference in the time evolution of the magnetization when $T_{ave}=20K$ and the anisotropy is varied between $K=0.1 \textrm{ mRyd}$ Fig.~\ref{fig:Figure7a} and $K=1 \textrm{ mRyd}$ Fig.~\ref{fig:Figure7b}. For low anisotropy constant the DW movement is much faster and the gradient dominates much more easily than for high anisotropy cases where the time evolution becomes less linear.}}
 \label{fig:Figure7}
\end{figure}

We have compared in Fig.~\ref{fig:Figure7} the change of behavior of two systems, for which the anisotropy strength is varied but all other parameters are kept the same. The systems with smaller anisotropy (Fig.~\ref{fig:Figure7a}) is much more susceptible to the thermal gradient than systems with higher anisotropy (Fig.~\ref{fig:Figure7b}). It is possible to observe that when the gradient is increased for the low anisotropy system, the time evolution of the magnetization  presents less fluctuations and it is much more linear than in the high anisotropy situation.

\begin{figure}[h]
 \centering
 \subfloat[$T_{ave}=20K$]{\label{fig:Figure8a}\includegraphics[width=0.79\columnwidth]{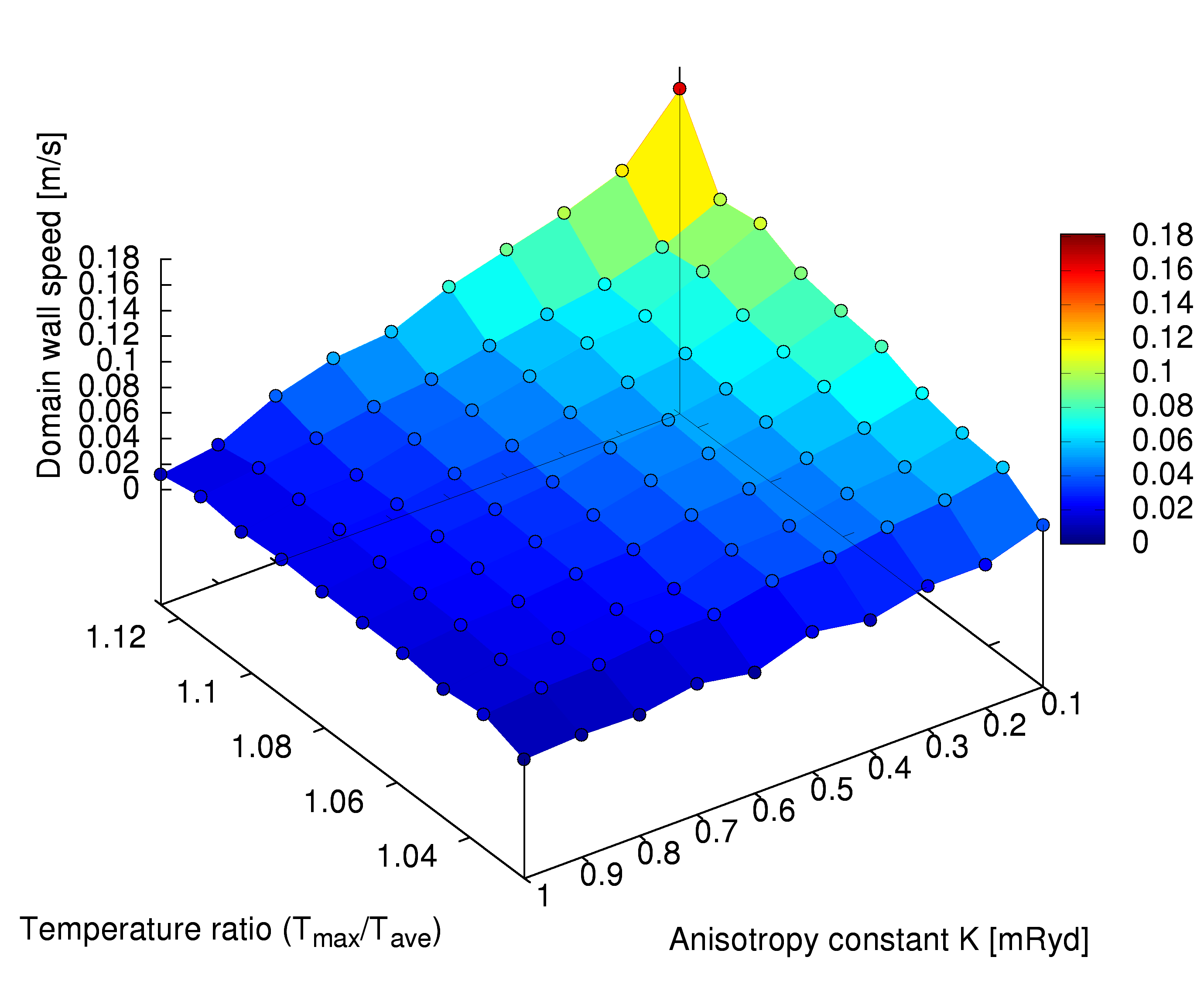}}\\\vspace{-2mm}
 \subfloat[$T_{ave}=60K$]{\label{fig:Figure8b}\includegraphics[width=0.79\columnwidth]{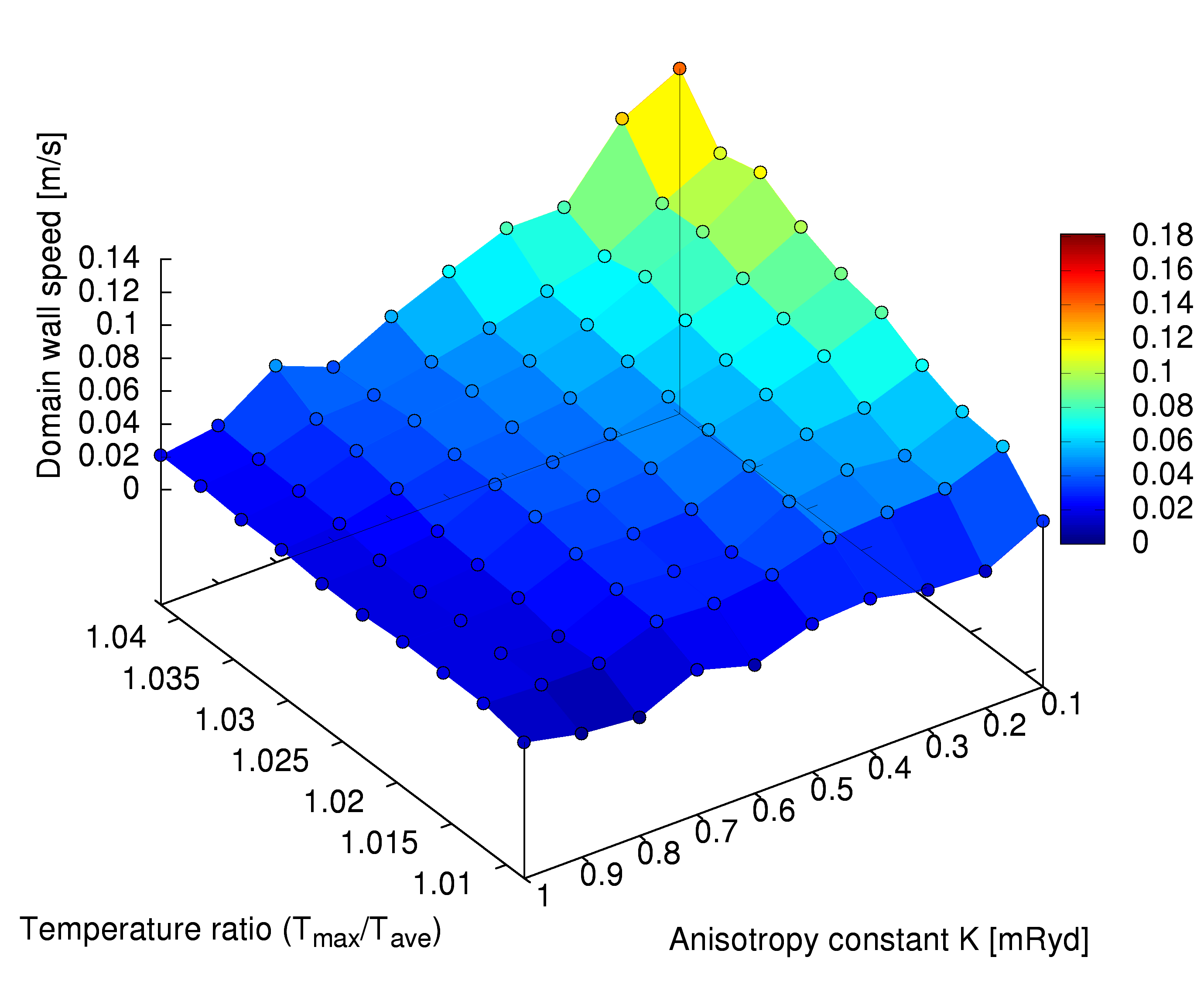}}\\\vspace{-2mm}
 \subfloat[$T_{ave}=100K$]{\label{fig:Figure8c}\includegraphics[width=0.79\columnwidth]{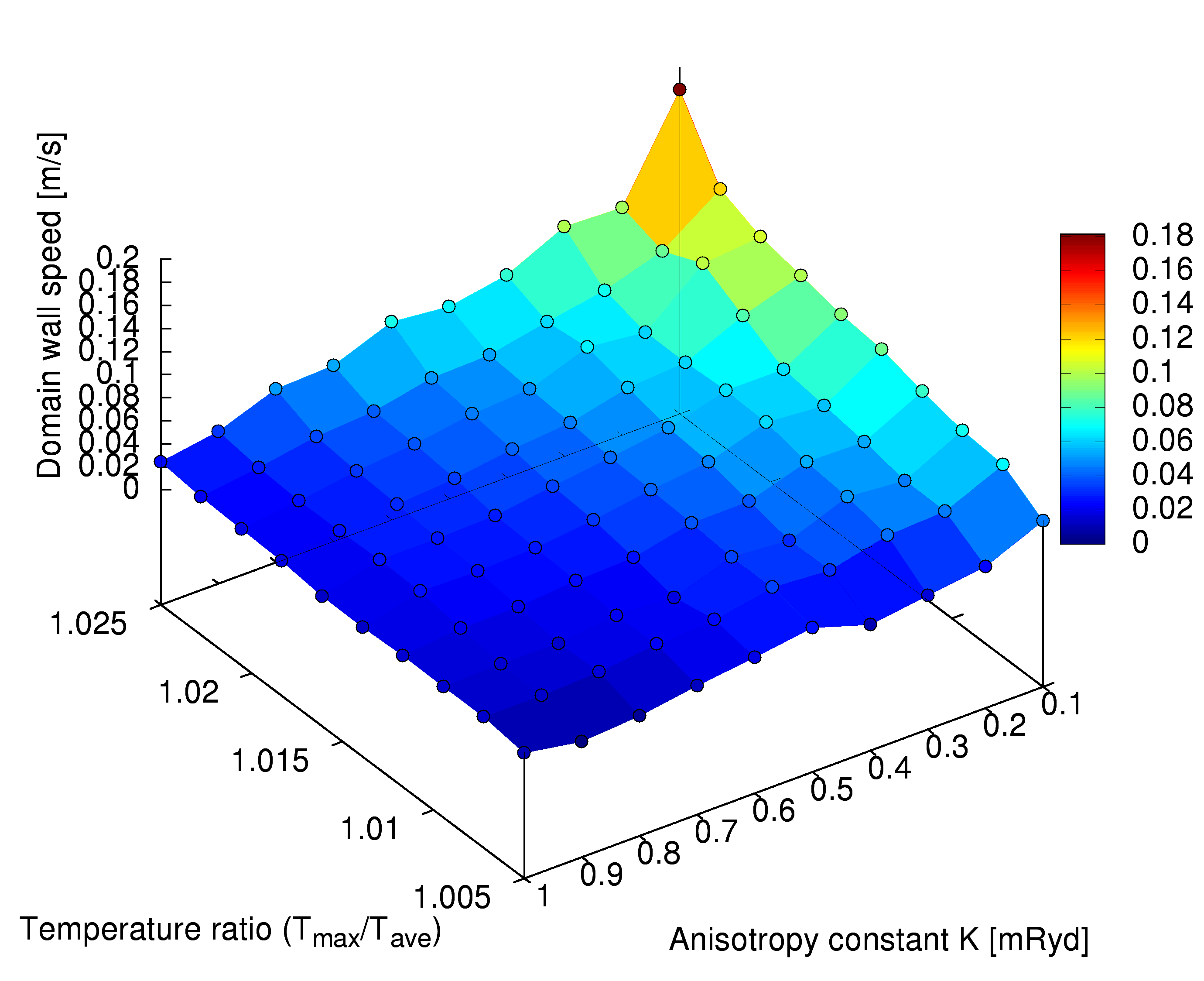}} 
 \caption{\small{Dependence of the domain wall speed on the anisotropy constant $K$ and the temperature ratio $\frac{T_{max}}{T_{ave}}$. As the anisotropy constant diminishes the DW speed that can be achieved with a given gradient for a given $T_{ave}$ increases.}}
 \label{fig:Figure8}
\end{figure}

A linear fit of the time evolution of the magnetization allows to plot the DW speed as a function of the temperature ratio $\frac{T_{max}}{T_{ave}}$ and the anisotropy constant $K$. It is possible to see how for a fixed $\frac{J}{K}$ as the temperature gradient increases the domain wall speed also increases (Fig.~\ref{fig:Figure8}). More importantly one can see that by making the anisotropy constant smaller (larger $\frac{J}{K}$) leads to larger DW widths and to a higher DW speed. This means that the wider the domain walls are, the higher the speeds they achieve under the influence of a thermal gradient. 

The question now is if we can identify why a broader DW moves faster under the influence of a given $\Delta T$. We can explain this behavior by examining the interactions present in the system. There are ferromagnetic exchange interactions which tend to align the spins in the same direction, while the uniaxial anisotropy wants to align the spins parallel to the easy magnetization axis. An atomically sharp DW minimizes the anisotropy energy, as all the spins are oriented parallel (or anti-parallel) to the anisotropy axis, but the exchange interaction is minimal if all the spins are have a minimal angle between them.

Thus to be able to make both interactions minimal at the same time, when the system is relaxed (it's evolved at a given constant temperature) the sum of all interactions will produce a region in which there is a smooth transition between parallel and anti-parallel orientations. To be able to minimize the energy as the anisotropy decreases, the DW width must increase as the exchange interaction will try to make as many spins parallel as possible.

When the thermal gradient is applied to the system, the magnons produced by the thermal fluctuations will interact with the spins in the domain wall. During these interactions, due to the angular momentum transfer, the spins will start changing their orientation (spins flip) causing the DW to move. If the domain wall is very narrow (i.e. the anisotropy is big) a large energy is needed to be able to flip one row of atoms. On the other hand, if the domain wall is wider (the case with low anisotropy) less energy is required to flip the spin orientation of atoms. Since the DW motion involves such spin-flip processes, it is to be expected that the low anisotropy case has a faster DW motion. This behavior is fully consistent with what was observed and discussed in Sec.~\ref{sec:few} and complements our arguments made there about the relationship between DW widths and their velocities in thermal gradients.

\section{Conclusions}
Extensive spin dynamics simulations have shown that it is possible to establish and move a domain wall in a atomically thin magnetic layer, here consisting of a monolayer of Fe on an W\hkl(110) surface, by applying a thermal gradient over the domain wall. The movement occurs here even though only thermally excited magnons are present and indirect effects from electron transport have not been considered.  The DW always prefers to move towards the hotter end of the sample, an observation that is consistent both with theories regarding magnon interactions with the DW and also a biased Brownian motion behavior of the DW which can be more relevant in the limit of very large magnetic anisotropies. 
\\
The  Fe / W\hkl(110) system shows very small domain wall speeds due to its  large anisotropy, which combined with its weak magnetic exchange coupling, determines a very narrow domain wall. Also it is possible to control the speed of the moving domain wall by changing the magnitude of the thermal gradient. The speed was found to be proportional to the thermal gradient. By considering model systems it has also been shown that if one lowers the anisotropy constant in a 2D system the domain wall width increases as expected, and the domain wall velocity increases as well. We find that it is a general trend that increasing the domain wall width/reducing the magnetic anisotropy increases the DW speed that can be obtained with a thermal gradient. 

The increased DW speed for lower magnetic anisotropy and larger DW widths, can be understood partially from the lower energy barriers associated to spin-flip processes when the DW moves. Another mechanism of explanation is that the DW can be seen as being moved by traveling magnons. A narrow DW width would decrease the velocity of the DW since it would provide a smaller cross section for the magnons to transfer angular momentum to the wall.
\\
The thermally induced domain wall motion reported here is reminiscent of the motion of domain walls in the presence of electrical currents and parametrically excited spin waves. Further studies, of both theoretical and experimental natures, comparing and combining these effects are anticipated to shed more light on this fascinating topic.

\section{Acknowledgments}
We gratefully acknowledge the European Research Council (ERC Project No. 247062-ASD), the Swedish Research Council (VR), the Knut and Alice Wallenberg Foundation, the Carl Tryggers Foundation, and the G\"oran Gustafsson Foundation for financial support. O.E. and A.B acknowledges eSSENCE. The computer simulations were performed on resources provided by the Swedish National Infrastructure for Computing (SNIC) at the National Supercomputer Centre (NSC) and High Performance Computing Center North (HPC2N). L.B and A.D acknowledges support from the Swedish e-Science Research Centre (SeRC)

\bibliographystyle{unsrt}
\bibliography{FeW_Thermal_Gradient}

\begin{thebibliography}{10}

\bibitem{seebeck_nat}
K.~Uchida, S.~Takahashi, K.~Harii, J.~Ieda, W.~Koshibae, K.~Ando, S.~Maekawa,
  and E.~Saitoh.
\newblock Observation of the spin seebeck effect.
\newblock {\em Nature}, 455(7214):778--781, 2008.

\bibitem{seebeck_semi}
C.~M. Jaworski, J.~Yang, S.~Mack, D.~D. Awschalom, J.~P. Heremans, and R.~C.
  Myers.
\newblock Observation of the spin-seebeck effect in a ferromagnetic
  semiconductor.
\newblock {\em Nature Materials}, 9:898--903, Sep 2010.

\bibitem{seebeck_in}
K.~Uchida, J.~Xiao, H.~Adachi, J.~Ohe, S.~Takahashi, J.~Ieda, T.~Ota,
  Y.~Kajiwara, H.~Umezawa, H.~Kawai, G.~E.~W. Bauer, S.~Maekawa, and E.~Saitoh.
\newblock {Spin Seebeck insulator}.
\newblock {\em Nature Materials}, 9(11):894--897, September 2010.

\bibitem{nowak}
D.~Hinzke and U.~Nowak.
\newblock Domain wall motion by the magnonic spin seebeck effect.
\newblock {\em Phys. Rev. Lett.}, 107:027205, Jul 2011.

\bibitem{phonon}
C.~M. Jaworski, J.~Yang, S.~Mack, D.~D. Awschalom, R.~C. Myers, and J.~P.
  Heremans.
\newblock Spin-seebeck effect: A phonon driven spin distribution.
\newblock {\em Phys. Rev. Lett.}, 106:186601, May 2011.

\bibitem{bauer2010}
Kjetil~M.D. Hals, Arne Brataas, and Gerrit~E.W. Bauer.
\newblock Thermopower and thermally induced domain wall motion in (ga, mn)as.
\newblock {\em Solid State Communications}, 150(11–12):461 -- 465, 2010.

\bibitem{Sampaio2012}
Tiago~S. Machado, Tatiana~G. Rappoport, and Luiz~C. Sampaio.
\newblock Vortex core magnetization dynamics induced by thermal excitation.
\newblock {\em Applied Physics Letters}, 100(11):112404, 2012.

\bibitem{Xia2010}
Zhe Yuan, Shuai Wang, and Ke~Xia.
\newblock Thermal spin-transfer torques on magnetic domain walls.
\newblock {\em Solid State Communications}, 150(11–12):548 -- 551, 2010.

\bibitem{Gerrit2010}
Gerrit E.~W. Bauer, Stefan Bretzel, Arne Brataas, and Yaroslav Tserkovnyak.
\newblock Nanoscale magnetic heat pumps and engines.
\newblock {\em Phys. Rev. B}, 81:024427, Jan 2010.

\bibitem{Stiles2008}
D.C. Ralph and M.D. Stiles.
\newblock Spin transfer torques.
\newblock {\em Journal of Magnetism and Magnetic Materials}, 320(7):1190 --
  1216, 2008.

\bibitem{Li2004}
S.~Zhang and Z.~Li.
\newblock Roles of nonequilibrium conduction electrons on the magnetization
  dynamics of ferromagnets.
\newblock {\em Phys. Rev. Lett.}, 93:127204, Sep 2004.

\bibitem{Niebala2007}
C.~Schieback, M.~Kläui, U.~Nowak, U.~Rüdiger, and P.~Nielaba.
\newblock Numerical investigation of spin-torque using the heisenberg model.
\newblock {\em The European Physical Journal B}, 59(4):429--433, 2007.

\bibitem{Wang2012}
Xi-guang Wang, Guang-hua Guo, Yao-zhuang Nie, Guang-fu Zhang, and Zhi-xiong Li.
\newblock Domain wall motion induced by the magnonic spin current.
\newblock {\em Phys. Rev. B}, 86:054445, Aug 2012.

\bibitem{Yan2011}
P.~Yan, X.~S. Wang, and X.~R. Wang.
\newblock All-magnonic spin-transfer torque and domain wall propagation.
\newblock {\em Phys. Rev. Lett.}, 107:177207, Oct 2011.

\bibitem{klaui1}
J.~H. Franken, P.~M\"{o}hrke, M.~Kl\"{a}ui, J.~Rhensius, L.~J. Heyderman, J.-U.
  Thiele, H.~J.~M. Swagten, U.~J. Gibson, and U.~R\"{u}diger.
\newblock Effects of combined current injection and laser irradiation on
  permalloy microwire switching.
\newblock {\em Applied Physics Letters}, 95(21):212502, 2009.

\bibitem{klaui2}
P.~Möhrke, J.~Rhensius, J.-U. Thiele, L.J. Heyderman, and M.~Kläui.
\newblock Tailoring laser-induced domain wall pinning.
\newblock {\em Solid State Communications}, 150(11–12):489 -- 491, 2010.

\bibitem{antropov}
V.~P. Antropov, M.~I. Katsnelson, B.~N. Harmon, M.~van Schilfgaarde, and
  D.~Kusnezov.
\newblock Spin dynamics in magnets: Equation of motion and finite temperature
  effects.
\newblock {\em Phys. Rev. B}, 54:1019--1035, Jul 1996.

\bibitem{skubic}
B~Skubic, J~Hellsvik, L~Nordström, and O~Eriksson.
\newblock A method for atomistic spin dynamics simulations: implementation and
  examples.
\newblock {\em Journal of Physics: Condensed Matter}, 20(31):315203, 2008.

\bibitem{hellsvik}
J.~Hellsvik.
\newblock {\em Atomistic Spin Dynamics, Theory and Applications.}
\newblock PhD thesis, Uppsala University, 2010.

\bibitem{anders}
Anders Bergman, Andrea Taroni, Lars Bergqvist, Johan Hellsvik, Bj\"orgvin
  Hj\"orvarsson, and Olle Eriksson.
\newblock Magnon softening in a ferromagnetic monolayer: A first-principles
  spin dynamics study.
\newblock {\em Phys. Rev. B}, 81:144416, Apr 2010.

\bibitem{Wiesendanger2001}
M.~Pratzer, H.~J. Elmers, M.~Bode, O.~Pietzsch, A.~Kubetzka, and
  R.~Wiesendanger.
\newblock Atomic-scale magnetic domain walls in quasi-one-dimensional fe
  nanostripes.
\newblock {\em Phys. Rev. Lett.}, 87:127201, Aug 2001.

\bibitem{Zakeri2010}
Kh. Zakeri, Y.~Zhang, J.~Prokop, T.-H. Chuang, N.~Sakr, W.~X. Tang, and
  J.~Kirschner.
\newblock Asymmetric spin-wave dispersion on fe(110): Direct evidence of the
  dzyaloshinskii-moriya interaction.
\newblock {\em Phys. Rev. Lett.}, 104:137203, Mar 2010.

\bibitem{Steiauf2005}
D.~Steiauf and M.~F\"ahnle.
\newblock Damping of spin dynamics in nanostructures: An \textit{ab initio}
  study.
\newblock {\em Phys. Rev. B}, 72:064450, Aug 2005.

\bibitem{VESTA}
Koichi Momma and Fujio Izumi.
\newblock {{\it VESTA3} for three-dimensional visualization of crystal,
  volumetric and morphology data}.
\newblock {\em Journal of Applied Crystallography}, 44(6):1272--1276, Dec 2011.

\bibitem{oguchi}
Takehiko Oguchi.
\newblock On the spin wave theory of bloch wall.
\newblock {\em Progress of Theoretical Physics}, 9(1):7--13, 1953.

\bibitem{Chantrell2008}
D.~Hinzke, N.~Kazantseva, U.~Nowak, O.~N. Mryasov, P.~Asselin, and R.~W.
  Chantrell.
\newblock Domain wall properties of fept: From bloch to linear walls.
\newblock {\em Phys. Rev. B}, 77:094407, Mar 2008.

\bibitem{Kazantseva}
N.~Kazantseva, R.~Wieser, and U.~Nowak.
\newblock Transition to linear domain walls in nanoconstrictions.
\newblock {\em Phys. Rev. Lett.}, 94:037206, Jan 2005.

\bibitem{Wees}
Gerrit E.~W. Bauer, Eiji Saitoh, and Bart~J. van Wees.
\newblock {Spin caloritronics}.
\newblock {\em Nature Materials}, 11(5):391--399, April 2012.

\bibitem{Torrejon2012}
J.~Torrejon, G.~Malinowski, M.~Pelloux, R.~Weil, A.~Thiaville, J.~Curiale,
  D.~Lacour, F.~Montaigne, and M.~Hehn.
\newblock Unidirectional thermal effects in current-induced domain wall motion.
\newblock {\em Phys. Rev. Lett.}, 109:106601, Sep 2012.

\end{thebibliography}

\end{document}